%% file: paper.tex
\documentclass[runningheads]{llncs}
\usepackage[T1]{fontenc}
\usepackage{graphicx}
\usepackage{xspace}
\usepackage{xcolor}
\usepackage{listings}
\usepackage{subcaption}
\usepackage{pifont}

\usepackage{draftwatermark}
\SetWatermarkText{PREPRINT}
\SetWatermarkScale{3}
\SetWatermarkLightness{0.93}

\usepackage{hyperref}
\usepackage[capitalize]{cleveref}
\newcommand\cpp{C\texttt{++}\xspace}
\newcommand\ctwozero{C\texttt{++}20\xspace}
\newcommand\ctwothree{C\texttt{++}23\xspace}
\newcommand{\cmark}{\textcolor[rgb]{0.0,0.5,0.0}{\ding{51}}\xspace} 
\newcommand{\xmark}{\textcolor[rgb]{0.5,0.0,0.0}{\ding{55}}\xspace} 
\newcommand{\astmark}{\textcolor[rgb]{0.7,0.3,0.0}{\ding{86}}\xspace} 

\crefname{figure}{Figure}{Figures}
\Crefname{figure}{Figure}{Figures}

\crefname{table}{Table}{Tables}
\Crefname{table}{Table}{Tables}
\definecolor{codegreen}{rgb}{0,0.6,0}
\definecolor{codepurple}{rgb}{0.58,0,0.82}
\lstdefinestyle{niceC++}{
    language=C++,
    frame=ltbr,
    rulecolor=\color{lightgray},
    commentstyle=\color{codegreen},
    keywordstyle=\color{blue},
    stringstyle=\color{red},
    basicstyle=\ttfamily,
    breaklines=false,
    columns=fullflexible,
    captionpos=b,
    upquote=true,
    keepspaces=true,
    numbers=none,
    showspaces=false,
    showstringspaces=false,
    showtabs=false,
    numberbychapter=false,
    morekeywords={nullptr, MPI_Datatype},
    literate={0} {{\color{codepurple}0}}2
             {1} {{\color{codepurple}1}}1
             {2} {{\color{codepurple}2}}1
             {3} {{\color{codepurple}3}}1
             {4} {{\color{codepurple}4}}1
             {__4} {4}1
             {5} {{\color{codepurple}5}}1
             {6} {{\color{codepurple}6}}1
             {7} {{\color{codepurple}7}}1
             {8} {{\color{codepurple}8}}1
             {9} {{\color{codepurple}9}}1,
    tabsize=2
}
\usepackage{color}

\begin{document}
\title{Layout-Agnostic MPI Abstraction for Distributed Computing in Modern C++}
\titlerunning{Layout-Agnostic MPI Abstraction}
\author{Jiří~Klepl\orcidID{0000-0002-2231-4073} \and
Martin~Kruliš\orcidID{0000-0002-0985-8949} \and
Matyáš~Brabec\orcidID{0009-0008-4470-2748}}
\authorrunning{J. Klepl et al.}
\institute{Charles University, Prague, Czech Republic\\
\email{\{klepl,krulis,brabec\}@d3s.mff.cuni.cz}}
\maketitle              
\begin{abstract}
Message Passing Interface (MPI) has been a well-established technology in the domain of distributed high-performance computing for several decades. However, one of its greatest drawbacks is a rather ancient pure-C interface. It lacks many useful features of modern languages (namely C++), like basic type-checking or support for generic code design. In this paper, we propose a novel abstraction for MPI, which we implemented as an extension of the C++ Noarr library. It follows Noarr paradigms (first-class layout and traversal abstraction) and offers layout-agnostic design of MPI applications. We also implemented a layout-agnostic distributed GEMM kernel as a case study to demonstrate the usability and syntax of the proposed abstraction. We show that the abstraction achieves performance comparable to the state-of-the-art MPI C++ bindings while allowing for a more flexible design of distributed applications.

\end{abstract}

\input{introduction}
\input{background}
\input{type-comparison}
\input{implementation}
\input{evaluation}
\input{conclusion}

\begin{credits}
\subsubsection{\ackname}
This paper was supported by the Johannes Amos Comenius Programme (P JAC, Natural and anthropogenic georisks) project CZ.02.01.01/00/22\_ 008/0004605, by Charles University institutional funding SVV (grant 260821), and Charles University Grant Agency (grant 269723).

\subsubsection{\discintname}
The authors have no competing interests to declare that are relevant to the content of this article.
\end{credits}

\bibliographystyle{splncs04}
\bibliography{bibliography}

\end{document}

%% file: introduction.tex
\section{Introduction}\label{sec:introduction}

Distributed computing is one of the essential ingredients of high-performance parallel computing. MPI (Message Passing Interface)~\cite{mpi41} represents one profound standard for this task, and it has many implementations, such as OpenMPI~\cite{openmpi50} or MPICH~\cite{mpich43}. Unfortunately, the MPI interface is rather old (considering the language development, especially in the past 25 years) and was designed on principles that are considered obsolete in modern \cpp, such as using \lstinline{void*} pointers for generically typed buffers. Furthermore, optimizing HPC algorithms often requires the utilization of multiple fine-tuned memory layouts for the data structures, which is somewhat clumsy in MPI since it requires tedious manual construction of MPI datatypes for individual layouts. This makes any support for layout-agnostic data structures or traversal-agnostic index spaces quite challenging.

\begin{figure}[htbp]
  \centering
  \begin{subfigure}{.25\linewidth}
      \centering
      \includegraphics[width=.9\linewidth]{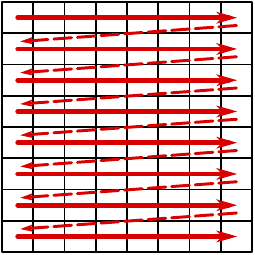}%
      \caption{row-major}%
      \label{fig:layout-row}
  \end{subfigure}%
  \begin{subfigure}{.25\linewidth}
      \centering
      \includegraphics[width=.9\linewidth]{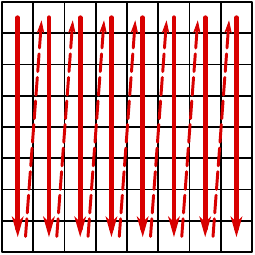}%
      \caption{col-major}%
      \label{fig:layout-column}
  \end{subfigure}%
  \begin{subfigure}{.25\linewidth}
      \centering
      \includegraphics[width=.9\linewidth]{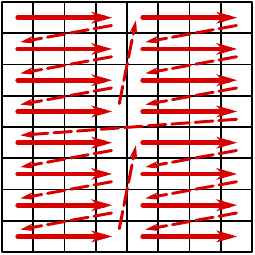}%
      \caption{tiled}%
      \label{fig:layout-tile}
  \end{subfigure}%
  \begin{subfigure}{.25\linewidth}
      \centering
      \includegraphics[width=.9\linewidth]{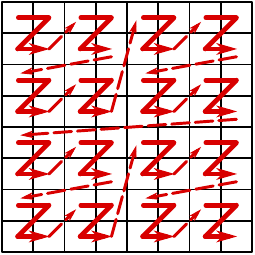}%
      \caption{z-curve}%
      \label{fig:layout-zcurve}
  \end{subfigure}%
  \caption{Examples of common matrix layouts}%
  \label{fig:matrix-layout}
  \vspace{-1em}
\end{figure}

Layout agnosticism~\cite{smelko2023astute} separates the logical index space of a data structure (like an abstraction or an interface) from its physical memory organization (implementation). A layout-agnostic data structure (e.g., a matrix) defines only its logical index space (such as the indices for the column $i$ and the row $j$) and provides an external mechanism (a layout definition) for mapping the logical indices to physical memory locations (offsets in a linear buffer). The programmer may choose different layout definitions to optimize the performance of the data structure for a given problem or specific hardware (memory) architecture while the algorithm that uses the data structure remains unchanged. A few examples of matrix memory layouts are depicted in \cref{fig:matrix-layout}.

The selected physical layout may affect the performance significantly. For instance, when multiplying two matrices $A\times B$ using a na\"{\i}ve $\mathcal{O}(N^3)$ algorithm, the first matrix $A$ is accessed in a row-wise pattern while the second $B$ is read column by column. Therefore, matrix $A$ could benefit from the row-major layout, while the column-major format would be better for $B$.

Unfortunately, selecting the best layout for a given problem/architecture is not always possible since a suboptimal data layout may be enforced by the API, and internal transformation of the data structure may be computationally demanding. However, a similar approach can be used to design traversal-agnostic algorithms~\cite{klepl2024abstractions} where iteration over an index space is abstracted, and the loops may be reordered, split into blocks, or even parallelized without changing the algorithm itself or the involved structures. Continuing with our matrix multiplication example, the three loops iterating over the $i,j,k$ dimensions (where $i,j$ represent an element in the output matrix and $k$ iterates over its dot product elements) can be split into blocks (in six nested loops) so the matrices are processed in a tiled manner, which improves data locality for most matrix layouts.

In recent years, there has been a growing interest in zero-cost overhead layout-agnostic data structures such as \lstinline{std::mdspan}~\cite{trott2022mdspan} included in the \ctwothree standard~\cite{WG21-N4950} (designed by Kokkos), Kokkos views~\cite{trott2021kokkos}, CuTe layouts~\cite{NVIDIA2025layout} by NVIDIA\@, or Noarr structures~\cite{klepl2024abstractions}. The traversal agnosticism has also been tackled from various standpoints. At the lowest level, the \cpp compiler attempts to provide automated loop optimizations based on the polyhedral model~\cite{pop2006graphite,grosser2012polly}. However, the compiler sometimes has a limited perspective of the semantics of the algorithm, so various extensions (like Loopy~\cite{namjoshi2016loopy}) were created that allowed the programmer to guide the compiler's loop transformations. An alternative approach is to use an abstraction layer for the data structure traversal. This may be achieved in a domain-specific language (DSL) such as Halide~\cite{ragan2012decoupling} or natively in \cpp using meta-template programming (e.g., with Noarr traversers~\cite{klepl2024abstractions}).

When designing a fine-tuned HPC solution, a programmer must pay attention to all aspects affecting efficiency, which involves combining low-level code optimizations, memory layout optimizations, parallelization, and distributed computing. Particularly in the context of MPI applications, the data distribution handled by MPI communication primitives needs to work in synergy with layout-agnostic and traversal-agnostic program design and with (potential) layout transformations to ensure optimal efficiency. Finally, the layout definitions should be propagated in the datatypes to ensure basic type safety, and these types need to be translated into MPI datatypes to ensure correct data transfers.

There are several libraries that provide MPI abstractions that may be considered for the layout-agnostic and traversal-agnostic approach. For example, Boost.MPI can handle direct transfers of STL vectors and serialization of many other data structures. Some works, such as MPP or KaMPIng~\cite{pellegrini2012lightweight,uhl2024kamping}, achieve similar goals with even better performance. Although these libraries offer some level of type safety and convenience when transferring simple containers, they are not designed for uses where the data needs to be also transformed to a different layout. KokkosComm~\cite{avans2024performance} allows the user to send or receive an arbitrary Kokkos view that represents an mdspan. However, its public interface assumes that the sender and receiver layouts are the same and unpredictably reorders the data for seemingly compatible Kokkos views, even with the same value type, rank, and dimension extents. This makes the library incompatible with the layout-agnostic approach without modifications to its interface.

We propose the Noarr-MPI library, a novel MPI abstraction that adopts principles and technologies from Noarr~\cite{klepl2024abstractions}, a \cpp library for managing data structure memory organization and traversal. We have implemented a prototype that handles basic wrapping and necessary interactions between Noarr and MPI libraries, automatic data transformations, initialization of MPI structures, and type checking. It also provides an MPI abstraction compatible with Noarr traversers.
The main contributions of the paper are:

\begin{itemize}
  \item We propose a novel layout-agnostic, type-checked abstraction for MPI, designed for modern \cpp and compatible with paradigms presented in the Noarr library.
  \item We provide an open-source prototype as an extension of the Noarr library\footnote{The implementation is available at \url{https://github.com/jiriklepl/noarr-mpi}}.
  \item We compare the features and performance of the Noarr-MPI library with state-of-the-art MPI abstractions, demonstrating the advantages of our approach.
  \item We demonstrate the proposed abstraction on a well-known matrix multiplication (GEMM) kernel, which serves as a practical example and provides a basis for comparison with other abstractions.
\end{itemize}

The paper is organized as follows.
\Cref{sec:background} introduces the Noarr library and related work.
\Cref{sec:layout-mapping} discusses mapping Noarr structures to MPI datatypes and the type transformation process.
The proposed implementation is detailed in \cref{sec:implementation}.
\Cref{sec:evaluation} evaluates the applicability of the proposed approach on a real-world example and presents the comparison with other abstractions. \Cref{sec:conclusion} concludes the paper.

%% file: background.tex
\section{Noarr Background}\label{sec:background}

Our work is built on top of the Noarr library~\cite{klepl2024abstractions}, so we start with a brief description of its main principles essential for our MPI abstraction.

\textbf{Noarr structures} are the core abstraction of the Noarr library. They represent a mapping from the index space with named dimensions to the linear memory space. Structures are assembled from \emph{proto-structure} objects that represent specific transformations of the mapping. For example, \lstinline{vector<'i'>(N)} introduces a dimension named \lstinline{'i'} with size \lstinline{N}, and \lstinline{into_blocks<'i','b'>(Ns)} splits dimension \lstinline{'i'} into two dimensions where \lstinline{'b'} is an index of a contiguous block of \lstinline{Ns} elements and \lstinline{'i'} represents individual elements within a block.

Multiple proto-structure objects can be combined using the \lstinline{^} operator---e.g., \lstinline{scalar<int>() ^ vector<'i'>(N) ^ vector<'j'>(M)} defines a 2D structure with dimensions named \lstinline{'i'} and \lstinline{'j'} of sizes \lstinline{N} and \lstinline{M}, respectively, and a scalar type of \lstinline{int}. This composition is hierarchical, meaning that the \lstinline{scalar<int>()} is the base structure further transformed by \lstinline{vector<'i'>(N)} and finally by \lstinline{vector<'j'>(M)}.

If \lstinline{'i'} represents the rows of a matrix and \lstinline{'j'} represents the columns, the \lstinline{scalar<int>() ^ vectors<'i', 'j'>(N, M)} (a shorthand for the above) structure represents the column-major layout, while for the row-major we simply swap the dimension names: \lstinline{scalar<int>() ^ vectors<'j', 'i'>(M, N)}.

Each Noarr structure has a \textbf{signature} that specifies the order of the dimensions in the structure. For the \lstinline{scalar<int>() ^ vectors<'i', 'j'>(N, M)} structure, the associated signature can be represented as $j \to i \to Int$. The signature becomes important when iterating over the structure, as it defines the default traversal order---in the example, dimension \lstinline{'j'} would be associated with the outermost loop.

The signature can be reordered using the \lstinline{hoist} proto-structure that shifts a dimension to the front (outermost) position. For example, \lstinline{hoist<'i'>} applied to the above structure would change the signature (and the default traversal order) to $i \to j \to Int$. The proto-structures change the signatures according to simple rewrite rules. For example, \lstinline{into_blocks<'i', 'b'>(Ns)} replaces the appearance of $i$ in the signature with $b \to i$, resulting in the signature $j \to b \to i \to Int$ if applied to the original structure. Besides the traversal order, the signatures also ensure the type safety of the transformations.

\textbf{Noarr bags} are smart pointers associating Noarr structures with data buffers. They simplify the use of structures in code where the layout-agnostic data structure is usually materialized in memory. A \lstinline{bag[idx<'i','j'>(i, j)]} expression accesses the element with index \lstinline{(i, j)} regardless of the actual memory layout. Bags are designed to support both owning and observing semantics.

\textbf{Noarr traversers} are objects representing a traversal order over an index space of a Noarr structure or multiple structures. When constructed, they follow a combination of the default traversal orders of the given structures (prioritizing from the left). Traversers can be transformed by applying \emph{proto-structures} to change the intended traversal order (similar to the construction of Noarr structures). For example, \lstinline{hoist} changes the iteration order of the dimensions, and \lstinline{span} restricts the iteration space along a given dimension. To ensure semantic correctness, proto-structures that change the physical layout (such as \lstinline{vector} that expands the memory layout) are not allowed for traversers; however, some of them have corresponding counterparts for traversers. For example, \lstinline{vector} can be replaced with \lstinline{bcast} that introduces a new dimension to the traversal (a new loop) without implying a change in the physical layout.

Given a set of Noarr bags and a Noarr traverser, the user can define a lambda function representing a computation and combine it with the traverser via the \lstinline{|} operator, as demonstrated in \cref{lst:noarr-mm}. The traverser applies the lambda function to each element of the (possibly transformed) iteration space represented by a \textbf{Noarr state} object containing the indices of a given point in the iteration space (such as \lstinline{idx<'i', 'j'>(i, j)}) that can be used to access the corresponding elements of the Noarr bags.

\begin{lstlisting}[basicstyle=\scriptsize\ttfamily,label={lst:noarr-mm},caption={Na\"{\i}ve matrix multiplication kernel in Noarr}]
// Allocate bags for the matrices with the given layouts:
auto C = bag(scalar<int>() ^ vector<'i'>(M) ^ vector<'j'>(N));
auto A = bag(scalar<int>() ^ vector<'i'>(M) ^ vector<'k'>(K));
auto B = bag(scalar<int>() ^ vector<'k'>(K) ^ vector<'j'>(N));

traverser(C) | [&](auto state) {  // C provides iterations space over (i,j)
  C[state] = 0;
  traverser(A, B) ^ fix(state) | [&](auto state) {  // iterating over k (i,j were fix-ed)
    C[state] += A[state] * B[state];  // [] applies relevant index sub-set of state
  };
};
\end{lstlisting}

Noarr also provides constructs to parallelize the traversal of selected dimension(s) based on existing parallel libraries like OpenMP (CPU) or CUDA (GPU).

\subsection{Related Work}\label{sec:related-work}

Various libraries and tools have been developed to make the use of MPI more convenient and less error-prone in the context of \cpp programming. We briefly discuss the most relevant libraries in the context of implementing layout-agnostic MPI abstractions. Most of these works have additional (orthogonal) goals, such as handling errors in MPI communication, or abstracting the non-blocking primitives to \cpp promises and futures.

As the oldest, still relevant work, the Boost.MPI~\cite{gregor2006boost} library provides a higher-level interface for \cpp STL containers and user-defined types in MPI communication by (de)serializing the data.
It introduces a form of layout agnosticism to MPI via a \emph{skeleton} object, which describes the layout of a particular data structure. Although the skeleton approach is quite generic (covering even linked lists or trees), its creation still involves costly (de)serialization. Furthermore, after creating the skeleton, the user has to ensure that the data structure is unchanged and that it is not paired with an incompatible skeleton (reducing the type safety of the communication).

The MPP library~\cite{pellegrini2012lightweight} improves on Boost.MPI by avoiding (de)serialization. It uses \lstinline{MPI_Type_create_struct} to define the layout of general data structures. It can also efficiently transform layouts of arbitrary \cpp containers by analyzing their type traits (querying the starting memory address of the elements, their size, and the number of elements).

Another improvement over Boost.MPI was introduced by the MPL header-only library~\cite{mpl}, which enables MPI communication of data structures by defining their \emph{layout}, which can be built (semi)automatically for trivial \cpp types and standard containers. The layouts map well to MPI datatypes; however, their construction for non-trivial layouts is cumbersome.

More recently, Demiralp~et~al.~\cite{demiralp2023c++} presented a modern \cpp library for MPI that uses compile-time reflection on \cpp types to generate appropriate MPI datatypes automatically. It introduced useful features such as exception-based error handling; however, its container support is limited to array-like contiguous containers (e.g., \lstinline{std::span} or \lstinline{std::vector}).

The KaMPIng library~\cite{uhl2024kamping} is the most recent major work that aims to introduce some layout agnosticism and type safety to MPI\@. It maps \cpp types to MPI datatypes at compile time via type introspection using template meta-programming. It supports standard \cpp containers and general trivially copyable types (which can be copied as contiguous sequences of bytes on homogeneous systems without needing to analyze their layout). However, it does not support non-contiguous containers, significantly limiting its usability in layout-agnostic applications.

The experimental KokkosComm library~\cite{avans2024performance} is a recent addition to the Kokkos ecosystem that extends the Kokkos library~\cite{trott2021kokkos} to support MPI communication. It is also the work that is closest to our Noarr-MPI abstraction in its general goal. Since Kokkos aims to maximize performance and portability, it offers some layout agnosticism via the \lstinline{Kokkos::View} abstraction. It stands out as the only related library that supports sending and receiving non-contiguous multidimensional data automatically without requiring any data preprocessing (such as serialization or data packing). The authors show that using the automatically generated MPI Datatypes in communication performs on par with hand-written datatypes. However, the KokkosComm library is still in its early stages and does not sufficiently support the necessary primitives to implement a fully layout-agnostic MPI abstraction---most notably, it does not fully support scatter and gather operations.

None of the existing MPI-abstracting libraries provide a layout-agnostic abstraction that could automatically handle MPI communication involving multiple data layouts with different compatible physical layouts (e.g., row-major and column-major matrices of the same size).
This is a significant limitation for use cases that call for a different layout of the data structures on the sending and receiving sides, which may easily happen if the layouts on the two sides are tuned independently to achieve the best performance. Using the existing libraries, the user has to manually define fitting MPI datatypes, which is tedious and error-prone, or resort to data packing and unpacking, which introduces additional overhead and is not transparent to the MPI backend. The Noarr-MPI library aims to eliminate this limitation by providing a layout-agnostic abstraction for MPI communication that can automatically handle the definition of the MPI datatypes based on the Noarr structures and traversers.

%% file: type-comparison.tex
\section{Noarr and MPI Layout Mapping}\label{sec:layout-mapping}

For successful Noarr-MPI binding, we need to define a mapping from Noarr structures to MPI datatypes. This mapping is non-trivial due to philosophical differences between the two libraries. On the other hand, datatypes both in MPI and in Noarr are constructed hierarchically from simple base types, giving us a common ground for the mapping.

\begin{lstlisting}[basicstyle=\scriptsize\ttfamily,label={lst:type-comparison},caption={Constructing an integer matrix in MPI and Noarr}]
auto row = scalar<int>() ^ vector<'n'>(n);
auto matrix = row ^ vector<'m'>(m);

/* MPI: */  MPI_Type_contiguous(n, MPI_INT, &row);
            MPI_Type_contiguous(m, row, &matrix);
\end{lstlisting}

\Cref{lst:type-comparison} illustrates two semantically identical definitions of a matrix layout in both libraries, showcasing their superficial similarities. The derived datatypes are strictly made of copies or sequences (so \lstinline{MPI_Type_contiguous} can be used). However, more complex data layouts (with tiling or slicing) will require more complex transformation.

\begin{lstlisting}[basicstyle=\scriptsize\ttfamily,label={lst:matrix-tiled},caption={Tiling the matrix structure}]
auto matrix_tiled = matrix ^ into_blocks<'m', 'M'>(M) ^ into_blocks<'n', 'N'>(N);

/* MPI: */  MPI_Type_contiguous(N, MPI_INT, &tileRow);
            MPI_Type_contiguous(n / N, tileRow, &row);
            MPI_Type_contiguous(M, row, &slice);
            MPI_Type_contiguous(m / M, slice, &matrix);
\end{lstlisting}

\Cref{lst:matrix-tiled} shows a tiled transformation of the \lstinline{matrix} type from the previous example that constructs a layout with the same underlying physical data but logically tiled into blocks of size ${M \times N}$ where each block is indexed by \lstinline{'M'} and \lstinline{'N'} (in Noarr). The contiguous blocks are specified separately to match the logical structure of the data. If a different logical ordering of the elements is required, the user has to use functions such as \lstinline{MPI_Type_create_hvector} and then carefully arrange the order of MPI calls and their arguments.

\subsection{Type Transformation}\label{sec:type-transformation}

Here, we present a general approach for transforming Noarr structures into MPI datatypes. The transformation is used internally by the Noarr-MPI bindings, but it can also be used directly by the user to create an RAII datatype handle that automatically releases the MPI datatype when it goes out of scope.

\begin{lstlisting}[basicstyle=\scriptsize\ttfamily]
auto matrix_tiled = matrix ^ into_blocks<'m', 'M'>(M) ^ into_blocks<'n', 'N'>(N);

auto datatype_handle = mpi_transform(matrix_tiled);
datatype_handle.commit();

MPI_Bcast(data, (MPI_Datatype)datatype_handle, root, comm);
\end{lstlisting}

The type transformation is defined recursively according to the dimension hierarchy encoded in the signature of the structure. For \lstinline{matrix_tiled}, the signature is $M \to m \to N \to n \to \texttt{Int}$. Generally, the signature is a tree with the outermost dimension in the root and the resulting scalar types in leaves. For each dimension, the process first constructs the MPI types of the nested dimensions and then uses them to construct the MPI type of the current dimension.

The datatype construction at each given dimension is usually equivalent to a single MPI call. The resulting MPI datatype has to match the traversal over the Noarr structure along the given dimension; namely, each $(\texttt{type}_i, \texttt{displacement}_i)$ pair in the resulting MPI datatype has to match the types and relative offsets of all sub-items $i$. Based on the types and displacements, an appropriate MPI call is selected:

\begin{enumerate}
  \item If the structure is stored contiguously and all types are the same (e.g., in the case of Noarr \lstinline{vector}), the \lstinline{MPI_Type_contiguous} can be used to construct the MPI datatype. This is the case for \lstinline{matrix_tiled} in \cref{lst:matrix-tiled}.
  \item If the structure is not stored contiguously but still follows constant strides, \lstinline{MPI_Type_create_hvector} can be used.
  \item If the offset of the first element is non-zero or the strides are not constant, we can use \lstinline{MPI_Type_create_hindexed} to specify offsets for individual items.
  \item If the structure is not uniform along the given dimension (i.e., $\texttt{type}_i \neq \texttt{type}_{i'}$ for some $i' > i$), the \lstinline{MPI_Type_create_struct} can be used to create irregular MPI structures.
\end{enumerate}

\begin{figure}[h]
  \vspace{-1.5em}
  \centering
  \includegraphics[width=0.8\linewidth]{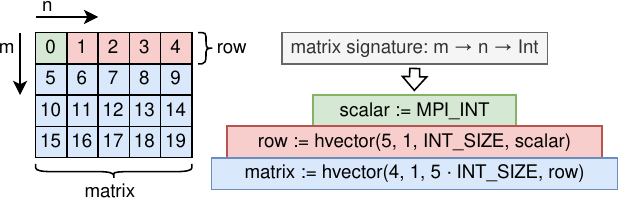}%
  \caption{Type transformation from Noarr structures to MPI datatypes}%
  \label{fig:type-transformation}
  \vspace{-1.25em}
\end{figure}

Noarr \lstinline{get_length} and \lstinline{get_size} functions are designed to determine the index ranges and the extents of the sub-structures. Additional trait-like functions, such as \lstinline{is_uniform_along}, \lstinline{stride_along}, and \lstinline{lower_bound_along}, can be used to select the appropriate MPI call.

\Cref{fig:type-transformation} showcases the construction from \cref{lst:type-comparison} disregarding the contiguity of the structure to illustrate the parametrization of the MPI calls. The block counts can be determined using \lstinline{get_length}, and the stride parameters can be computed using \lstinline{stride_along}.

\subsection{Layout-agnostic Type Transformation}\label{sec:agnostic-type-transformation}

Our main objective is to enable the transfer of data structures with identical index spaces but different physical layouts. This will require data transformation during the transfer, which can be achieved seamlessly with the MPI communication primitives as long as the MPI datatypes are constructed properly.

To build on the matrix example, let us consider a broadcast operation where the source and destination structures are column-major and row-major matrices, respectively. Na\"{\i}vely, we might construct a contiguous MPI datatype for each matrix, which would be efficient; however, it would also corrupt the data on transfer due to differing orderings of elements between the two structures, which is the case for frameworks like KokkosComm~\cite{avans2024performance}. We need to ensure that the MPI datatypes are compatible and that they correctly represent the given data structures at the same time.

There are multiple ways to construct compatible MPI datatypes; perhaps the simplest one would be to reorder the signatures to a canonical form (e.g., lexicographically ordered dimensions). However, this can result in suboptimal transfers (the performance depends on how the layouts are described, not just on the layouts themselves), and it is impractical to affect efficiency by a mere choice of dimension names. To avoid this, a traverser can be provided to define the intended hierarchy of dimensions during the transformation via its signature. Since the traverser has the same index space as its Noarr structure parameters, we can construct it from the list of the structures involved in the communication to ensure correctness. Furthermore, the user may transform the traverser for the purpose of performance tuning (e.g., apply \lstinline{noarr::hoist} to reorder the dimensions).

Constructing the MPI datatype for \lstinline{matrix} according to the default traverser of the transposed matrix differs from the construction in \cref{fig:type-transformation} only in the dimension hierarchy (switching the order of \lstinline{hvector} calls), which changes the order of the elements in the resulting MPI datatype. However, unlike the previous example, in this case, the \lstinline{hvector} calls cannot be replaced with \lstinline{contiguous}. The resulting MPI datatype is compatible with \lstinline{MPI_Type_contiguous} over the memory space of the transposed matrix, enabling transfers between the two structures without data corruption.

%% file: implementation.tex
\section{MPI Interface Abstraction}\label{sec:implementation}

Our MPI interface extends the Noarr traverser abstraction (for iterating over an index space) into an \emph{MPI traverser} that combines a regular traverser with an MPI communicator, which enables a seamless extension of existing Noarr algorithms.

\subsection{MPI Traverser}\label{sec:mpi-traverser}

MPI traversers abstract iteration over an index space distributed over multiple ranks (MPI processes) in the same group (MPI communicator). They are inspired by Noarr CUDA traversers~\cite{klepl2024abstractions}, which abstract work distribution over a CUDA thread grid (running on a GPU).

The construction of MPI traversers is shown in \cref{lst:mpi-traverser}. The constructor selects one dimension as the \emph{ranking dimension} (\lstinline{'r'}), and its index is bound to the MPI rank. The size of the ranking dimension must match the size of the communicator, and it is set automatically to this size if not specified explicitly. The resulting MPI traverser can be used as a regular Noarr traverser and in place of an MPI communicator in Noarr-MPI bindings (communication operations).

\begin{lstlisting}[basicstyle=\scriptsize\ttfamily,label={lst:mpi-traverser},caption={Defining an MPI traverser over the dimension \lstinline{'r'}}]
auto trav = traverser(matrix ^ into_blocks<'m', 'r', 's'>());
auto MPITrav = mpi_traverser<'r'>(trav, MPI_COMM_WORLD);
auto slice = bag(scalar<float>() ^ vectors_like<'s', 'm'>(MPITrav));

if (mpi_get_comm_rank(MPITrav) == 0) {
  // Using `MPITrav` as a Noarr traverser (applying a lambda for execution)
  MPITrav | [&](auto state) { slice[state] = init(); };
}
// Using `MPITrav` as an MPI communicator for broadcast operation
broadcast(slice, MPITrav, 0);
\end{lstlisting}

In \cref{lst:mpi-traverser}, the matrix sized ${m \times n}$ is sliced along the \lstinline{'m'} dimension into $r$ sub-matrices ${{m / r} \times n}$ where $r$ is determined automatically from the communicator size. The root node initializes the slice and then broadcasts it to all nodes in the communicator associated with the traverser.

\subsection{Collective Communication}\label{sec:collective-communication}

Simple type transformation from Noarr structures to MPI datatypes is sufficient for broadcasts and point-to-point communication that use the same logical layout for both the sending and the receiving ranks. However, more complex operations like scatter and gather involve data structures with not only different physical layouts but also different logical layouts---the index space of one input structure being a subspace of the other input structure. The scatter and gather are representative collective operations for our transformations, and since they are symmetric, we discuss only the scatter operation in detail.

A na\"{\i}ve approach to scattering of general layouts would involve costly data (de)serialization; however, this can be avoided with the Noarr library and the type transformations described in \cref{sec:agnostic-type-transformation}, making the operation layout-agnostic thanks to constructing the MPI datatypes from the Noarr structures according to the logical ordering of elements dictated by the same traverser.

After constructing the MPI datatypes for the two input structures, the scatter operation uses the ranking dimension of the MPI traverser to determine the correct displacements of the sub-structures and scatters the data accordingly. Since the ranking dimension is bound to the MPI rank, it maps to the tiles of the source structure.

\begin{lstlisting}[basicstyle=\scriptsize\ttfamily,label={lst:scatter},caption={Scatter operation using an MPI traverser}]
auto matrix = bag(matrix_tiled, buffer);

// We need to fix one block dimension (M=__4) and associate r with blocking dimensions MxN
auto trav = traverser(matrix) ^ set_length<'M'>(4) ^ merge_blocks<'M', 'N', 'r'>();
auto MPITrav = mpi_traverser<'r'>(trav, MPI_COMM_WORLD);

// Specifying the layout of the tile (with matching index space)
auto tile = bag(scalar<float>() ^ vectors_like<'m', 'n'>(MPITrav));

if (mpi_get_comm_rank(MPITrav) == 0) {
  traverser(matrix) ^ set_length(MPITrav) | [&](auto state) { matrix[state] = init(); };
}
scatter(matrix, tile, MPITrav, 0);

MPITrav | [&](auto state) { modify(tile[state]); };
\end{lstlisting}

\Cref{lst:scatter} uses the tiled matrix from \cref{lst:matrix-tiled} and scatters it across all ranks where the tiles are processed by the \lstinline{modify} function. The key aspect here is that we wish to divide the matrix into tiles so that each rank processes one tile. However, dimensions \lstinline{'M'} and \lstinline{'N'} are both open, so we need to fix one of them to a specific value (in this case, \lstinline{'M'} is fixed to 4) so the remaining dimension can be set based on the number of ranks $N = r/M$ (which is ensured by the \lstinline{merge_blocks} association). The ${M \times N}$ grid of tiles is distributed so that MPI ranks $0\ldots{N-1}$ compute row of tiles with \lstinline{'M'} index $0$, ranks $N\ldots{{2N}-1}$ compute row where \lstinline{'M'} is $1$, and so on. Currently, our implementation requires that $M$ is selected as a divisor of $r$, and original matrix dimensions ($m,n$) must be divisible by $M$ and $N$, respectively.

Perhaps the greatest benefit is that the layout of the tiles (\lstinline{tile} structure) can be defined independently of the layout of the original matrix. The scatter operation will perform the necessary transformation that converts the layout, allowing the layout-agnostic design of the computations on the tiles to fully exploit the memory architecture and caches of the target node.

The proposed abstraction also ensures the type safety of the operation. The index space of the distributed structure has to be a subspace of the root structure index space, and the difference in the index spaces has to be covered by the dimension bound to the MPI communicator. This eliminates the possibility of a mismatch between the input structures and simplifies the API\@.

%% file: evaluation.tex
\section{Evaluation}\label{sec:evaluation}

We evaluate the proposed Noarr-MPI library by comparing it with selected MPI abstractions in terms of the supported features, ease of use, and performance of the generated code. In the evaluation, we include the MPI libraries introduced in \Cref{sec:related-work}, which demonstrate at least some overlap with our main objectives (layout agnosticism and type safety): \textbf{MPI} (using standard MPI 4.0 interface) with \cpp mdspans, \textbf{Boost.MPI}~\cite{gregor2006boost} with \cpp mdspans, \textbf{MPP}~\cite{pellegrini2012lightweight}, \textbf{MPL}~\cite{mpl}, \textbf{KokkosComm}~\cite{avans2024performance}, and \textbf{KaMPIng}~\cite{uhl2024kamping}. Our proposed solution is denoted \textbf{Noarr-MPI}.

For the approaches that support a sufficient level of layout agnosticism, we perform a case study of implementing a distributed general matrix multiplication (GEMM) kernel as defined in the Polybench/C benchmark suite~\cite{pouchet2016polybench}, with each MPI rank computing one sub-matrix of the output matrix by multiplying the corresponding sub-matrices of the input matrices. We discuss the necessary steps to implement the GEMM kernel in \cref{sec:programming-effort-discussion} and evaluate the performance of the implementations in \cref{sec:performance-evaluation}.

\subsection{Feature Comparison}\label{sec:feature-comparison}

The comparison of the individual libraries focuses mainly on the features related to the layout agnosticism.

\begin{description}
  \item[Automatic transformations] of layouts if the source and destination ranks use a different layout datatype specification. This is a crucial feature for layout agnosticism as it allows for defining the layout of each data structure independently without specifying explicit data transformations.
  \item[Non-contiguous layouts] --- i.e., support for data structures that are not laid in one compact block of memory. This is especially essential for scatter and gather operations where many sub-structures will be non-contiguous.
  \item[Mdspan-like] representation for the data structures involved in the communication. It allows the user to reason about layouts in terms of the logical order of the elements in the data structure while the library automatically handles the construction of the MPI datatypes.
  \item[Seamless] libraries avoid unnecessary data packing or serialization when transferring the data structures. This is not a hard requisite, but it may improve efficiency.
  \item[Type safety] of a library indicates that it should fail to compile the code if the source and destination data structures have incompatible index spaces.
  \item[Scatter/gather operations] of multi-dimensional data structures are a crucial feature for layout agnosticism. Without this feature, the user has to specify a linearized view of the data, which intricately depends on the data layout and introduces unnecessary complexity and potential errors.
\end{description}

\Cref{tab:feature-comparison} summarizes the features of the evaluated libraries. The \astmark mark for the MPI support of auto-transformations denotes that these transformations are technically possible, but they require significant additional effort. The datatypes must be constructed imperatively, not automatically inferred, like in Noarr-MPI.

\begin{table}[htbp]
  \vspace{-2em}
  \centering
  \caption{Comparison of the MPI abstractions}%
  \label{tab:feature-comparison}
  \begin{tabular}{|l|c|c|c|c|c|c|c|}
    \hline
     & Noarr-MPI & native MPI & Boost.MPI & ~MPP~ & ~MPL~ & Kokkos & KaMPIng \\
    \hline
    Auto-transforms & \cmark & \astmark & \xmark & \xmark & \xmark & \xmark & \xmark \\
    \hline
    Non-contiguous & \cmark & \cmark & \cmark & \cmark & \cmark & \cmark & \xmark \\
    \hline
    Mdspan-like & \cmark & \xmark & \xmark & \xmark & \xmark & \cmark & \xmark \\
    \hline
    Seamless & \cmark & \cmark & \xmark & \cmark & \cmark & \cmark & \cmark \\
    \hline
    Type-safety & \cmark & \xmark & \cmark & \cmark & \cmark & \cmark & \cmark \\
    \hline
    Scatter/gather  & \cmark & \cmark & \xmark & \xmark & \cmark & \xmark & \xmark \\
    \hline
  \end{tabular}
  \vspace{-2em}
\end{table}

\subsection{Programming Effort Discussion}\label{sec:programming-effort-discussion}

Since conducting a comprehensive user study is beyond the scope of this paper, we provide a discussion of the necessary programming effort to implement the GEMM kernel using the libraries that support a sufficient level of layout agnosticism as described in \cref{sec:feature-comparison} (supporting more than half of the discussed features). Using text-based or syntax-based metrics for this comparison would be misleading as the libraries differ significantly in their approach, and any practical layout-agnostic use of the libraries would be accompanied by support abstractions that would significantly reduce the amount of necessary boilerplate code. Instead, we focus on the necessary steps and required information that the user has to provide to implement the GEMM kernel using the evaluated libraries, disregarding the specific syntax or verbosity that could be hidden behind a shorthand abstraction.

An ideal tunable and generalizable implementation of the kernel would require only the \emph{definition of the data layout} of each matrix and its privatized tile on a given worker process, the \emph{scattering strategy}, the implementation of \emph{the computation}, and some specification of \emph{data dependencies} (in all the evaluated libraries, this is done via explicit calls representing MPI communication). For each of these general steps, we discuss the specific sub-steps that are needed to implement the GEMM kernel using the evaluated libraries.

\subsubsection{Data Layout Definition}\label{sec:data-layout-definition}

When using the Boost.MPI library or the standard MPI interface, we have used the \lstinline{mdspan} abstraction designed by the Kokkos group and adopted in the \ctwothree standard~\cite{trott2022mdspan} to define the data layout of each matrix and its privatized tiles. Using mdspans is sufficiently expressive and convenient for layout agnosticism. They are compatible with the libraries and do not introduce any unnecessary overhead.
For the KokkosComm library, we use Kokkos views, which are very similar to the \lstinline{mdspan} abstraction and the Kokkos library uses them natively.

Both of these approaches require the user to define the index ranges of each sub-matrix tile separately (and match them correctly), which creates an unnecessary source of potential errors. Noarr structures in the Noarr-MPI approach allow us to specify the index ranges collectively for all tiles. Additionally, the library can automatically deduce the ranges of the tiles (e.g., based on the number of MPI ranks).

\subsubsection{Scattering Strategy}\label{sec:scattering-strategy}

Each library has its own way of defining how to scatter and gather the data. The Noarr-MPI library infers the correct MPI datatypes from the input and output data structures (\cref{sec:agnostic-type-transformation}). The inferred datatypes can be further tuned (for optimizing transformations) by reordering the logical order (hierarchy) of the dimensions of the given traverser (\cref{sec:mpi-traverser}). Noarr-MPI also allows the user to easily specify the mapping of the tiles to the worker threads.

In the approach that uses just the standard MPI interface, we build the MPI datatypes by iteratively using the \lstinline{MPI_Type_create_hvector} function with the extent and the stride of each dimension of the given mdspan from the rightmost to the leftmost (we have experimentally verified that this is the most efficient method for this use case). This generalizes well to an arbitrary number of dimensions; however, this approach may be suboptimal for other use cases.

The KokkosComm library internally uses the same approach; however, its API does not support scattering and gathering. They offer non-blocking receive and send operations that can be used to emulate the scatter and gather operations with very little overhead but at much higher programming effort. Since the library does not consider layout agnosticism, its receive and send interface is unusable for our purposes (it works inconsistently for contiguous and non-contiguous data layouts). However, the library does publicly expose the necessary building blocks for layout-agnostic send and receive operations.

The Boost.MPI library does not support the scatter and gather operations natively for this use case, so we have opted to try two different approaches: the first one is to define wrappers that specify how the items of a mdspan should be serialized and deserialized via a simple iteration in the logical order of the elements and scatter this serialized data. The other approach is to use input and output archives in non-blocking send and receive operations; however, the second approach performed consistently worse, so it is not included in the evaluation. Using skeletons in the Boost.MPI is too cumbersome for this use case and is not included in the evaluation.

\subsubsection{Computation Kernel Implementation}\label{sec:computation-kernel-implementation}

The computation kernel is implemented similarly across all libraries and is rather straightforward (almost verbatim copy of the GEMM loop nest from the PolyBench/C benchmark).

The only outlier is the Noarr-MPI library, which uses Noarr traversers. As analyzed in our previous work~\cite{klepl2024abstractions}, the traverser abstraction significantly reduces the amount of explicit indexation, which can prevent potential errors and makes the code more flexible for possible tuning. In this aspect, the Noarr library is similar to \lstinline{std::ranges} in \ctwozero, which also allows for a more declarative design of algorithms. Even with a slight syntactical difference between the Noarr traversers and an explicit loop nest, all libraries follow the same conceptual pattern of the computation kernel and specifying it via a loop nest can be directly 1-to-1 mapped to Noarr traversers.

\subsection{Performance Evaluation}\label{sec:performance-evaluation}

In the performance evaluation, we measure only the code that would be executed repeatedly in a real-world application (i.e., the \lstinline{scatter}, \lstinline{compute}, and \lstinline{gather} operations). We do not include the initialization of the MPI environment, allocation of the buffers, data validation, or cleanup operations. To ensure a fair comparison, all implementations allocate the data the same way, and all libraries use views with non-owning semantics and without any additional data movements except those necessary for scattering and gathering the data. Most importantly, all implementations perform the same exact computation in the same order and on the same data layouts. We have also ensured that all implementations follow the same data distribution strategy (i.e., send/receive the same data on each given rank). The details and all measured data can be found in our replication package\footnote{The replication package is available at \url{https://github.com/jiriklepl/noarr-mpi}}.

The evaluation was performed on a Slurm-managed cluster with 8 nodes interconnected via InfiniBand FDR (56 Gb/s), each with two sockets comprising $2\times 32$ virtual cores. All GEMM implementations were compiled with GCC 14.2 and OpenMPI 5 using the \texttt{Release} configuration as defined by CMake 3.27. The evaluation used the various datasets from the Polybench/C benchmark, which were modified to be divisible by the number of MPI ranks; however, we visualize only the two opposite ends of the spectrum, \texttt{MINI} and \texttt{EXTRALARGE} (all dataset results are available in the replication package), which show interesting performance trends. For the \texttt{MINI} dataset, all dimensions are $64$, and for the \texttt{EXTRALARGE} dataset, the dimensions are $n_i = 2048, n_j = 2560, n_k = 1408$. The performance is measured in seconds by a wall clock, and each visualized result is the average and standard deviation of 100 runs of the distributed GEMM kernel.

\begin{figure*}[htbp]
  \vspace{-1.5em}
  \centering
  \includegraphics[width=\linewidth]{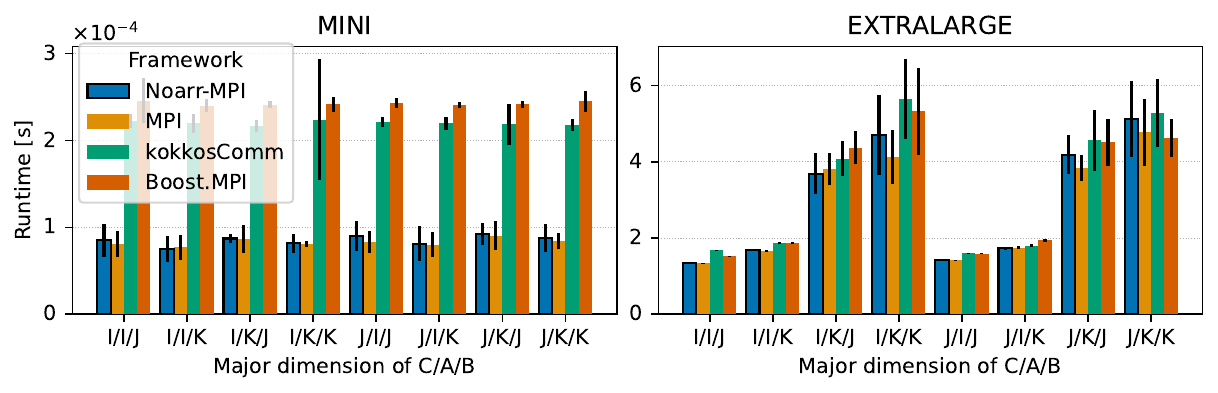}
  \caption{Performance evaluation of the evaluated libraries (columns show mean runtime in seconds, error bars show standard deviation)}%
  \label{fig:compare}
  \vspace{-1.25em}
\end{figure*}

\Cref{fig:compare} shows the results of the performance evaluation. The individual configurations (labeled on the x-axis) represent the major dimensions of the privatized tiles used in GEMM for the \texttt{C}, \texttt{A}, and \texttt{B} matrices written as \texttt{\emph{C}/\emph{A}/\emph{B}} (using dimension names from \cref{lst:noarr-mm}). For example, the \texttt{I/I/J} configuration denotes that the \texttt{C} and \texttt{A} matrix tiles are laid out in a row-major order and the \texttt{B} matrix tiles are laid out in a column-major order. All implementations passed the validation checks.

The measurements performed on the \texttt{MINI} dataset show that the \texttt{Noarr-MPI} library performs on par with the standard MPI interface for \cpp, while the other approaches are significantly slower. The overheads on this dataset are dominated by the non-computational parts of the code (i.e., communication latency, datatype construction, and data movements during scattering). The \texttt{SMALL} and \texttt{MEDIUM} datasets show a similar trend as the \texttt{MINI} dataset, but the differences between the libraries are less pronounced.

In the case of the \texttt{EXTRALARGE} dataset, computational time significantly dominates the communication overhead. For most configurations, the \texttt{Noarr-MPI} library performs on par or slightly better than the other libraries. The only significant outliers that show a performance drop when using the \texttt{Noarr-MPI} library are the \texttt{I/K/K} and \texttt{J/K/K} configurations, where the serialization strategy of the \texttt{Boost.MPI} implementation outperforms the other libraries, which defer the layout transformation to the MPI backend. However, for all four most efficient configurations, the \texttt{Noarr-MPI} library performs the best. The \texttt{LARGE} dataset shows a similar trend, but \texttt{Noarr-MPI} (and other libraries) do not experience the performance drop on the four less efficient configurations.

The comparison shows that the proposed Noarr-MPI library can achieve performance comparable to the state-of-the-art MPI \cpp bindings while allowing for a more flexible design of distributed applications and thus improving the performance without expending the programming effort necessary to move the data correctly. Except for a few outliers, its performance is equivalent to hand-written MPI code using the \cpp mdspan abstraction for computation.

%% file: conclusion.tex
\section{Conclusion}\label{sec:conclusion}

Noarr-MPI is a novel \cpp abstraction for MPI that introduces a layout-agnostic design to distributed computing. It builds on top of the Noarr library and closely follows its paradigms. We have used the abstraction to implement an MPI version of the GEMM kernel to demonstrate its applicability and performance compared to the state-of-the-art MPI \cpp bindings and a hand-written implementation using the standard MPI interface. Our evaluation shows that the proposed abstraction achieves performance comparable to the state-of-the-art MPI \cpp bindings while allowing for a more flexible design of distributed applications. Furthermore, it enables programmers to design distributed applications without expending effort on data layout transformations or complicated MPI datatype definitions. The abstraction is also compatible with Noarr traversers, allowing for seamless cooperation with CUDA and OpenMP traversers.

%% file: paper.bbl
\begin{thebibliography}{10}
\providecommand{\url}[1]{\texttt{#1}}
\providecommand{\urlprefix}{URL }
\providecommand{\doi}[1]{https://doi.org/#1}

\bibitem{avans2024performance}
Avans, C.N., Ciesko, J., Pearson, C., Suggs, E.D., Olivier, S.L., Skjellum, A.: Performance insights into supporting kokkos views in the kokkoscomm mpi library. In: 2024 IEEE International Conference on Cluster Computing Workshops (CLUSTER Workshops). pp. 186--187. IEEE (2024)

\bibitem{mpl}
Bauke, H.: {MPL - A message passing library}. \url{https://github.com/rabauke/mpl} (2014-2015)

\bibitem{demiralp2023c++}
Demiralp, A.C., Martin, P., Sakic, N., Kr{\"u}ger, M., Gerrits, T.: {A C++ 20 Interface for MPI 4.0}. arXiv preprint arXiv:2306.11840  (2023)

\bibitem{gregor2006boost}
Gregor, D., Troyer, M.: Boost. mpi. MPI, November  (2006)

\bibitem{grosser2012polly}
Grosser, T., Groesslinger, A., Lengauer, C.: Polly: performing polyhedral optimizations on a low-level intermediate representation. Parallel Processing Letters p. 1250010 (2012)

\bibitem{WG21-N4950}
{ISO/IEC JTC1/SC22/WG21}: Working draft, standard for programming language c++. Tech. Rep. N4950 (May 2023), \url{https://open-std.org/jtc1/sc22/wg21/docs/papers/2023/n4950.pdf}, revises: N4944

\bibitem{klepl2024abstractions}
Klepl, J., {\v{S}}melko, A., Rozsypal, L., Kruli{\v{s}}, M.: {Abstractions for C++ code optimizations in parallel high-performance applications}. Parallel Computing p. 103096 (2024)

\bibitem{mpi41}
{Message Passing Interface Forum}: {MPI}: A Message-Passing Interface Standard Version 4.1 (nov 2023), \url{https://www.mpi-forum.org/docs/mpi-4.1/mpi41-report.pdf}

\bibitem{mpich43}
{MPICH Team}: {MPICH v4.3.0 Manpages} (2025), \url{https://www.mpich.org/static/docs/v4.3.0/}

\bibitem{namjoshi2016loopy}
Namjoshi, K.S., Singhania, N.: Loopy: Programmable and formally verified loop transformations. In: International Static Analysis Symposium. pp. 383--402. Springer (2016)

\bibitem{NVIDIA2025layout}
NVIDIA: {CuTe Layout Documentation} (2025), \url{https://github.com/NVIDIA/cutlass/blob/main/media/docs/cpp/cute/01_layout.md}, documentation on CuTe layout from the NVIDIA CUTLASS repository

\bibitem{pellegrini2012lightweight}
Pellegrini, S., Prodan, R., Fahringer, T.: {A lightweight C++ interface to MPI}. In: 2012 20th Euromicro International Conference on Parallel, Distributed and Network-based Processing. pp. 3--10. IEEE (2012)

\bibitem{pop2006graphite}
Pop, S., Cohen, A., Bastoul, C., Girbal, S., Silber, G.A., Vasilache, N.: {GRAPHITE: Polyhedral analyses and optimizations for GCC}. In: proceedings of the 2006 GCC developers summit. pp. 90--91. Citeseer (2006)

\bibitem{pouchet2016polybench}
Pouchet, L.N., Yuki, T.: {PolyBench/C 4.2.1} (May 2016), \url{https://sourceforge.net/projects/polybench}

\bibitem{ragan2012decoupling}
Ragan-Kelley, J., Adams, A., Paris, S., Levoy, M., Amarasinghe, S., Durand, F.: Decoupling algorithms from schedules for easy optimization of image processing pipelines. ACM Transactions on Graphics (TOG) pp. 1--12 (2012)

\bibitem{smelko2023astute}
{\v{S}}melko, A., Kruli{\v{s}}, M., Kratochv{\'\i}l, M., Klepl, J., Mayer, J., {\v{S}}im{\r{u}}nek, P.: Astute approach to handling memory layouts of regular data structures. In: Algorithms and Architectures for Parallel Processing: 22nd International Conference, ICA3PP 2022, Copenhagen, Denmark, October 10--12, 2022, Proceedings. pp. 507--528. Springer (2023)

\bibitem{openmpi50}
{The Open MPI Community}: {Open MPI v5.0.x — Open MPI 5.0.x documentation} (2025), \url{https://docs.open-mpi.org/en/v5.0.x/}

\bibitem{trott2022mdspan}
Trott, C., Hollman, D., Lebrun-Grandie, D., Hoemmen, M., Sunderland, D., Edwards, H.C., Adelstein~Lelbach, B., Bianco, M., Sander, B., Iliopoulos, A., Michopoulos, J., Liber, N.: {MDSPAN}. Tech. Rep. P0009r18, WG21 (Jul 2022), \url{https://www.open-std.org/jtc1/sc22/wg21/docs/papers/2022/p0009r18.html}

\bibitem{trott2021kokkos}
Trott, C.R., Lebrun-Grandie, D., Arndt, D., Ciesko, J., Dang, V., Ellingwood, N., Gayatri, R., Harvey, E., Hollman, D.S., Ibanez, D., et~al.: Kokkos 3: Programming model extensions for the exascale era. IEEE Transactions on Parallel and Distributed Systems pp. 805--817 (2021)

\bibitem{uhl2024kamping}
Uhl, T.N., Schimek, M., H{\"u}bner, L., Hespe, D., Kurpicz, F., Seemaier, D., Stelz, C., Sanders, P.: {KaMPIng: Flexible and (Near) Zero-Overhead C++ Bindings for MPI}. In: SC24: International Conference for High Performance Computing, Networking, Storage and Analysis. pp. 1--21. IEEE (2024)

\end{thebibliography}
